\documentclass[AMA,STIX1COL,12pt,doublespace]{WileyNJD-v2}
\usepackage{moreverb}

\articletype{Article Type}%

\received{<day> <Month>, <year>}
\revised{<day> <Month>, <year>}
\accepted{<day> <Month>, <year>}

\usepackage{amsmath}
\usepackage{threeparttable}
\usepackage[hyphenbreaks]{breakurl}

\begin{document}
	
	\title{Estimands and Inference in Cluster-Randomized Vaccine Trials\protect}
	
	\author[1]{Kayla W. Kilpatrick*}
	
	\author[1]{Michael G. Hudgens}
	
	\author[2]{M. Elizabeth Halloran}
	
	\authormark{Kilpatrick \textsc{et al.}}

	\address[1]{\orgdiv{Department of Biostatistics}, \orgname{University of North Carolina}, \orgaddress{\city{Chapel Hill}, \state{NC}, \country{U.S.A.}}}
	
	\address[2]{\orgdiv{Center for Inference and Dynamics of Infectious Diseases, Vaccine and Infectious Disease Division}, \orgname{Fred Hutchinson Cancer Research Center}, \orgaddress{\city{Seattle}, \state{WA}, \country{U.S.A.}}}

	\corres{*Kayla W. Kilpatrick, \orgdiv{Department of Biostatistics}, \orgname{University of North Carolina}, \orgaddress{\city{Chapel Hill}, \state{NC}, \country{U.S.A.}}. \email{kkilpat@live.unc.edu}}
	
	
	\abstract[Abstract]{Cluster-randomized trials are often conducted to assess vaccine effects. Defining estimands of interest before conducting a trial is integral to the alignment between a study's objectives and the data to be collected and analyzed. This paper considers estimands and estimators for overall, indirect, and total vaccine effects in trials where clusters of individuals are randomized to vaccine or control. The scenario is considered where individuals self-select whether to participate in the trial and the outcome of interest is measured on all individuals in each cluster. Unlike the overall, indirect, and total effects, the direct effect of vaccination is shown in general not to be estimable without further assumptions, such as no unmeasured confounding. An illustrative example motivated by a cluster-randomized typhoid vaccine trial is provided.}
	
	

\maketitle

\section{Introduction}\label{sec1}
Vaccines are integral to combating a variety of infectious diseases. Quantifying a vaccine's effects is vital to determining its benefits, which can then guide public health policies aimed at reducing the burden of disease. Cluster-randomized trials are often conducted to quantify the effects of a treatment or intervention such as a vaccine. In cluster-randomized trials, individuals are grouped together based on certain characteristics (e.g., neighborhood of residence), and the entire cluster is randomized to treatment or control. The process of randomization ensures that the treatment and control groups are exchangeable. Cluster-randomization is useful when it is impractical or infeasible to randomize at the individual level \citep{Halloran2010}. Comparisons between randomized clusters can be used to assess the overall impact of an intervention on the population, which is particularly important in settings where an intervention may have indirect (or spillover) effects \citep{Hayes2000}. For example, in the infectious disease setting, whether one individual is vaccinated could affect the outcome of another individual. Moulton et al. \cite{Moulton2001} describe a cluster-randomized trial in the White Mountain Apache Reservation and the Navajo Nation wherein approximately 9000 infants within 38 clusters were randomized by cluster to the vaccine of interest (\textit{Streptococcus pneumoniae} conjugate vaccine) or control (a meningococcal C conjugate vaccine). Diallo et al. \cite{Diallo2019} present a cluster-randomized trial of an inactivated influenza vaccine in Senegal in which approximately 7800 enrolled, age-eligible children within 20 clusters were randomized by cluster to the influenza vaccine or control (an inactivated polio vaccine). Sur et al. \cite{Sur2009} describe a cluster-randomized trial of a typhoid vaccine in India, with approximately 38000 individuals within 80 clusters randomized by cluster to the typhoid vaccine or control (hepatitis A vaccine).

Because the cluster-randomized trial is a common study design for evaluating vaccine effects, it is important to carefully define the estimands, i.e., parameters of interest, in these trials. Careful definition of the effects of interest prior to the study can aid in study planning and can ensure that the study's goals are achieved \cite{Leuchs2015}. Recently, there has been increased interest in defining estimands in clinical trials. The International Council on Harmonization (ICH) has published a draft addendum to the E9 guidelines detailing the use of estimands in clinical trials and is currently in the process of refining and finalizing the addendum \citep{ICH}. This addendum aims to describe the necessity of defining the target estimand before the design and analysis of trials to avoid misalignment of the trial goals and the data, as well as to ensure that estimation of the estimand is possible without relying upon dubious assumptions \citep{Mehrotra2016}.

Leuchs et al.\cite{Leuchs2015}, Koch and Wiener\cite{Koch2016}, Permutt\cite{Permutt2016}, and Phillips et al.\cite{Phillips2017} discuss examples of estimands of interest in regulatory clinical trials. Target estimands specifically for cluster-randomized trials have been previously considered for certain designs. Wu et al.\cite{Wu2014} consider estimands for matched-pair cluster-randomized trials. Hudgens and Halloran \cite{Hudgens2008} consider estimands of the direct, indirect, total, and overall effects of treatment assuming a two-stage randomization scheme. In this design, clusters are randomly assigned to a treatment allocation program, and individuals within the clusters are randomly assigned to treatment based on the cluster-level assignment. In some cluster-randomized trials, individuals may not comply with their randomization assignment or may choose not to participate in the study\cite{Moulton2001, Sur2009, Surcholera, senegal}. Frangakis et al. \cite{Frangakis2002clustered} consider clustered encouragement designs, which allow noncompliance, where individuals belong to one of three principal strata: always-takers, compliers, and never-takers. Kang and Keele\cite{Kang2018} also consider cluster-randomized trials with noncompliance. Like Frangakis et al.\cite{Frangakis2002clustered}, they consider the setting where there are the three principal strata mentioned above, and also the special case where there are no always-takers. Even for this special case, they show the total and indirect (spillover) effects are not identified because principal strata membership is unknown for some individuals.

In this paper, we consider cluster-randomized vaccine trials where individuals choose whether or not to participate in the trial. As illustrated by the examples described above, it is common in cluster-randomized vaccine trials for the control to be another vaccine which is not expected to affect the outcome of interest. For simplicity, below the control vaccine will sometimes be referred to just as a control. Here we consider the particular case where a control vaccine is employed and individuals are blinded, i.e., unaware whether their cluster is randomly assigned to the vaccine of interest or to the control vaccine. In this setting, it is reasonable to assume individual participation behavior is unaffected by randomization, such that there are only two principal strata: always participators and never participators. Thus, our setting is similar to the special case considered by Kang and Keele\cite{Kang2018}. However, because it is assumed an individual will participate or not in the trial regardless of randomization assignment, principal strata membership is known for all individuals; this allows for identification and estimation of overall, total and indirect effects.

Sur et al.\cite{Sur2009} provides a motivating example of a cluster-randomized vaccine trial where individuals self-select whether to participate. In this trial, clusters of individuals were randomized to either a typhoid vaccine or a control vaccine (for hepatitis A). The presence of a control allowed study blinding, so individuals in the clusters did not know which assignment their cluster received. While some individuals chose not to participate in the trial, outcome data was collected on all individuals. This allows inference about different effects of the vaccine, as described below.

The outline of the remainder of this paper is as follows. In Section \ref{Methods}, notation, estimands, estimators, and effects of interest are described. In Section \ref{MotivatingExample}, the Sur et al. \cite{Sur2009} cluster-randomized typhoid vaccine trial is considered. Finally, Section \ref{Conclusion} concludes with a discussion.

\section{Methods} \label{Methods}
\subsection{Notation and Potential Outcomes}
Consider a cluster-randomized vaccine trial with $n$ clusters (or groups) of individuals where each cluster is randomly assigned to vaccine or control. For $i=1,\ldots, n$, let $A_i=1$ if cluster $i$ is assigned to vaccine and $A_i=0$ otherwise. Let $Y_i^{a=1}$ denote the potential outcome if cluster $i$ is assigned vaccine, and let $Y_i^{a=0}$ denote the potential outcome if cluster $i$ is assigned control. For example, $Y_i^{a=1}$ could denote the proportion of individuals in cluster $i$ who would develop typhoid within one year after randomization if, possibly counter to fact, cluster $i$ were assigned to vaccine. For now, we leave the particular outcome associated with $Y_i^a$ unspecified. Different specifications of $Y_i^a$ will correspond to different vaccine effects, as described below. Let $Y_i$ denote the observed outcome for cluster $i$, such that $Y_i=Y_i^{a=1}A_i+Y_i^{a=0}(1-A_i)$. Below, the subscript $i$ is sometimes dropped for notational convenience.

In cluster-randomized vaccine trials, one individual's vaccination status may affect another individual's outcome, that is, there may be ``interference" between individuals\cite{Cox1958}. For instance, if one individual receives a typhoid vaccine, this could affect whether or not another individual develops typhoid. Throughout this paper, it is assumed that there is no interference between individuals in different clusters, i.e., there is ``partial interference"\cite{Sobel2006}. Under this assumption, the outcome $Y_i$ for cluster $i$ depends only on the treatment assigned to cluster $i$. No assumption is made regarding the form of interference within clusters.

\subsection{Estimands and Estimators}
Vaccine effects, i.e., the causal effects of vaccination, can be defined by contrasts in the expected values of the potential outcomes $Y^{a=1}$ and $Y^{a=0}$. Assuming the $n$ clusters in the trial are randomly sampled from an infinite super-population of clusters, the average treatment (vaccine) effect is generally defined by
\begin{equation} \label{eqestimand2}
\theta=E[Y^{a=1}]-E[Y^{a=0}]
\end{equation} 
where $E[X]$ denotes the expected value of $X$ in the super-population of clusters. In words, (\ref{eqestimand2}) is the difference in the average outcome in the super-population when a cluster receives $a=1$ compared to when a cluster receives $a=0$. Alternatively, the $n$ clusters could be considered the finite population of interest and $E[X]$ defined instead to be $n^{-1} \sum_{i=1}^n X_i$. The super-population perspective is adopted in this paper, but similar considerations to those provided here apply if the finite population approach is utilized instead. Likewise, estimands other than (\ref{eqestimand2}) could be considered. For example, for binary $Y$, the risk ratio $E[Y^{a=1}]/E[Y^{a=0}] = \Pr[Y^{a=1}=1]/\Pr[Y^{a=0}=1]$ might be of greater interest than the risk difference (\ref{eqestimand2}). More generally, causal effects can by defined by $g(E[Y^{a=1}],E[Y^{a=0}])$ for some contrast function $g(x,y)$ where $g(x,x)=0$; e.g., $g(x,y)=x-y$ corresponds to (\ref{eqestimand2}). Below, estimands of the form (\ref{eqestimand2}) are described, but similar considerations apply for other contrasts.

A few aspects of defining causal effects bear mentioning. First, causal effects are typically defined by contrasts in expected values of the potential outcomes over the same set of units\cite{Rubin1974, Frangakis2002}. In many settings, the unit is defined to be an individual; for example, a unit could be a participant in a randomized controlled trial. Here, we consider the clusters to be the units since randomization is at the cluster level. Note that contrasts in average potential outcomes between different sets of units do not have a causal interpretation. For example, suppose a cluster-randomized vaccine trial is conducted in schools, where students within the same school constitute the clusters. A comparison of the average $Y^{a=1}$ among clusters (schools) in rural areas to the average $Y^{a=0}$ among clusters in urban areas is not a causal effect. Also note that causal effects are contrasts in the expected value of the {\it same} outcome under different counterfactual scenarios. Contrasts in different outcomes are not causal effects. For example, a comparison of the average incidence of typhoid when clusters receive vaccine with the average incidence of cholera when clusters receive control would not be a causal effect. We will revisit this point below when discussing direct effects.

The average treatment effect can be estimated by the difference in sample means:
\begin{equation}\label{eqestimator}
\hat{\theta}=\frac{\sum_{i=1}^n Y_iI(A_i=1)}{\sum_{i=1}^n I(A_i=1)}-\frac{\sum_{i=1}^n Y_iI(A_i=0)}{\sum_{i=1}^n I(A_i=0)}
\end{equation}  
This estimator is consistent and unbiased under commonly used randomization schemes, such as a completely randomized experiment where the number of clusters assigned vaccine (treatment) is fixed \citep{Miratrix2013, Imbens2015, Athey2017}. The standard error of $\hat{\theta}$ can be estimated and 95\% Wald confidence intervals can be constructed in the usual manner for the difference in means. Equivalently, (\ref{eqestimator}) can be obtained by computing the least squares estimate of the slope parameter of simple linear regression of $Y$ on $A$.  A generally more precise estimator can be obtained by regressing $Y$ on $A$ and $Z$ where $Z$ is some vector of baseline covariates. For simplicity, only estimators of the form (\ref{eqestimator}) are considered below; see Tsiatis et al.\cite{tsiatis2008} for further discussion on using baseline covariates to improve efficiency.


\subsection{Overall, Indirect, and Total Effects}\label{OE}
In this section, the general approach above is used to define estimands and estimators of the overall, indirect, and total effects. The outcome of interest will depend on the context of the vaccine trial, such as the infection or pathogen of interest, the target population, and so forth. Here, the outcome of interest is generically referred to as disease.

The overall effect compares the average disease outcome among all individuals when a cluster is assigned vaccine versus when a cluster is assigned control. This quantity may be the most relevant to public health policy because all individuals within clusters are used in the comparison. As it is likely that populations of interest will include a mixture of individuals who would and who would not choose to be vaccinated, the overall effect may be valuable for public health officials and policy makers in assessing the overall impact of a vaccine at the population level.

The overall effect estimand and estimator can be defined in terms of individual level outcomes as follows. Let $m_i$ denote the number of individuals in cluster $i$. For individual $j$ in cluster $i$, let $Y_{ij}=1$ if individual $j$ develops disease, and let $Y_{ij}=0$ otherwise. Let $Y_{ij}^{a=1}$ indicate the outcome that would have been observed for individual $j$ if cluster $i$ is randomized to vaccine, and define $Y_{ij}^{a=0}$ analogously for control, such that $Y_{ij} = Y_{ij}^{a=1} A_i + Y_{ij}^{a=0} (1-A_i)$. For the overall effect, the estimand (\ref{eqestimand2}) can be expressed in terms of individual potential outcomes by defining $Y_i^{a=1}=\sum_{j=1}^{m_i} Y_{ij}^{a=1}/m_i$,  and $Y_i^{a=0}=\sum_{j=1}^{m_i} Y_{ij}^{a=0}/m_i$ for cluster $i$. The overall effect estimator can likewise be expressed in terms of the observed individual-level outcomes by letting $Y_i=\sum_{j=1}^{m_i} Y_{ij}/m_i$.

The indirect effect quantifies the effect of vaccination on individuals who chose not to participate in the trial and, therefore, have no chance of receiving the vaccine. This effect is defined as a contrast in the average outcomes among non-participants when their cluster does or does not receive vaccine\cite{Halloran1991}. Because the indirect effect is defined only among individuals who never receive the vaccine, this effect (if present) is solely due to interference. Thus, indirect effects are a type of spillover or peer effect\cite{Sobel2006}.  Quantifying indirect effects may be of interest from a public health policy perspective because vaccinating some, but not all, individuals within a cluster can still provide benefits to those who are unable or choose not to be vaccinated.

Like the overall effect, the indirect effect estimand and estimator can be defined in terms of individual level outcomes. To do so, first define the potential outcome $S_{ij}^{a=1}$ where $S_{ij}^{a=1}=1$ if individual $j$ in cluster $i$ would choose to participate in the trial if, possibly counter to fact, cluster $i$ were randomized to vaccine and $S_{ij}^{a=1}=0$ otherwise. Define $S_{ij}^{a=0}$ analogously. Denote the observed participation outcome for individual $j$ in cluster $i$ by $S_{ij}$, such that $S_{ij} = S_{ij}^{a=1}A_i + S_{ij}^{a=0}(1-A_i)$. Assume $S_{ij}^{a=1}=S_{ij}^{a=0}$, i.e., an individual's decision to participate is not affected by whether their cluster is assigned vaccine or control. This assumption may be reasonable in cluster-randomized trials where individuals are blinded, such as the typhoid vaccine trial described in Section \ref{MotivatingExample}, because in such settings, randomization is not expected to have an effect on an individual's decision to participate in the trial. As mentioned in the Introduction, Frangakis et al.\cite{Frangakis2002} and Kang and Keele\cite{Kang2018} utilize the principal stratification framework when considering non-compliance in cluster-randomized trials. Under the assumption $S_{ij}^{a=1}=S_{ij}^{a=0}$, all individuals belong to one of two principal strata: always participators, i.e., individuals where $S_{ij}^{a=1}=S_{ij}^{a=0}=1$; and never participators, i.e., individuals where $S_{ij}^{a=1}=S_{ij}^{a=0}=0$. Fortunately, unlike the setting considered by Kang and Keele, here the principal strata membership of each individual can be inferred directly from the observed data because $S_{ij} = S_{ij}^{a=1} = S_{ij}^{a=0}$.


The indirect effect is the effect of vaccine in the non-participator principal stratum. The indirect effect has the general form (\ref{eqestimand2}), with $Y_i^{a=1}$ now defined to be $\left\{\sum_{j=1}^{m_i} Y_{ij}^{a=1}I(S_{ij}^{a=1}=0)\right\}/\left\{\sum_{j=1}^{m_i}I(S_{ij}^{a=1}=0)\right\}$ and $Y_i^{a=0}$ defined to be $\left\{\sum_{j=1}^{m_i} Y_{ij}^{a=0}I(S_{ij}^{a=0}=0)\right\}/\left\{\sum_{j=1}^{m_i}I(S_{ij}^{a=0}=0)\right\}$. This estimand compares the average disease outcome among non-participators when a cluster is assigned vaccine versus when a cluster is assigned control. Similarly, the indirect effect estimator can be expressed by (\ref{eqestimator}) with $Y_i$ defined to be $\left\{\sum_{j=1}^{m_i} Y_{ij}I(S_{ij}=0)\right\}/\left\{\sum_{j=1}^{m_i}I(S_{ij}=0)\right\}$.

The total effect measures the effect of treatment in the always participator principal stratum. Because always participators receive the vaccine if and only if their cluster is assigned vaccine, the total effect encompasses both the individual effect of receiving the vaccine as well as the effect of other individuals in the cluster being vaccinated. The total effect estimand and estimator have the same form as the indirect effect estimand and estimator described above, but with $S_{ij}^{a=1}=0$ replaced by $S_{ij}^{a=1}=1$, $S_{ij}^{a=0}=0$ replaced by $S_{ij}^{a=0}=1$, and $S_{ij}=0$ replaced by $S_{ij}=1$. The total effect quantifies the difference in the average disease outcome among always participators when a cluster is assigned vaccine versus when a cluster is assigned control. The total effect is often the effect of primary interest in this type of trial. An illustration of the overall, indirect, and total effects is given in Figure 1. 


There are a few special cases of note. In the scenario where all individuals in the population are willing to participate in trials (i.e., there are no non-participators), the indirect effect is not well-defined, and the total and overall effects are equivalent.  In some trials, only a subset of individuals may be eligible to be randomized for vaccination. For example, in Sur et al.\cite{Sur2009}, individuals were eligible if they were at least two years of age, were not pregnant or lactating, and did not have an elevated temperature when the vaccine was given. Indirect effects, analogous to that defined above for non-participators, can be defined and estimated in these individuals if their outcome of interest is measured.

\begin{figure}[h]
	\centering
	\includegraphics[width=0.5\linewidth]{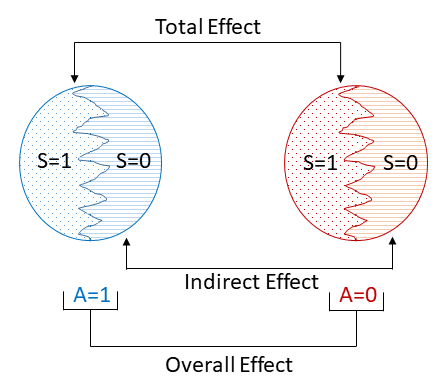}
	\caption{Cluster counterfactual comparisons. The left circle represents a cluster if, possibly counter to fact, assigned to vaccine ($A=1$). The right circle represents a cluster if, possibly counter to fact, assigned to control ($A=0$). Within each circle, $S$ indicates which individuals chose to participate in the study ($S=1$ indicates participation, $S=0$ otherwise). The overall, indirect, and total effects are contrasts in average potential outcomes over different sets of individuals within the clusters.}
	\label{fig:sur-cluster-randomized-trial-design-figurev3}
\end{figure}

\subsection{Direct Effect}
The overall, indirect, and total effects each describe an effect of treatment (vaccination) which is at least partially due to interference, if present. The effect of treatment that is not attributable to interference may also be of interest. Such an effect is sometimes referred to as a direct effect. This section describes why it is not possible in general to estimate the direct effect of vaccination in a cluster-randomized trial with self-selection of participation without additional assumptions, such as no unmeasured confounding. Informally, the direct effect compares the average outcome when an individual is vaccinated to the average outcome when an individual is not vaccinated, holding fixed the proportion of other individuals vaccinated\cite{Halloran1991}. Several formal definitions of the direct effect estimand have been proposed; e.g., see Hudgens and Halloran\cite{Hudgens2008},VanderWeele and Tchetgen Tchetgen\cite{VanderWeele2011}, Liu et al. \cite{Liu2016}, Eck et al.\cite{Eck2018} and S\"{a}vje et al.\cite{Savje2018}.


Naively, it might be tempting to estimate the direct effect by comparing the proportion of vaccinated individuals with disease to the proportion of unvaccinated individuals with disease in clusters assigned to vaccine, i.e., by
\begin{equation}\label{DEestimator}
\frac{1}{\sum_{i=1}^n I(A_i=1)}\sum_{i=1}^n \frac{\sum_{j=1}^{m_i} Y_{ij}I(S_{ij}=1)}{\sum_{j=1}^{m_i}I(S_{ij}=1)}I(A_i=1)-\frac{1}{\sum_{i=1}^n I(A_i=1)}\sum_{i=1}^n \frac{\sum_{j=1}^{m_i} Y_{ij}I(S_{ij}=0)}{\sum_{j=1}^{m_i}I(S_{ij}=0)}I(A_i=1).
\end{equation}
\noindent
 However, (\ref{DEestimator}) converges in probability to 
\begin{equation}\label{DEestimand}
E[Y^{a=1}]-E[\tilde{Y}^{a=1}]
\end{equation}  
where $Y_i^{a=1}=\left\{\sum_{j=1}^{m_i} Y_{ij}^{a=1}I(S_{ij}^{a=1}=1)\right\}/\left\{\sum_{j=1}^{m_i}I(S_{ij}^{a=1}=1)\right\}$ and $\tilde{Y}_i^{a=1}=\left\{\sum_{j=1}^{m_i} Y_{ij}^{a=1}I(S_{ij}^{a=1}=0)\right\}/\left\{\sum_{j=1}^{m_i}I(S_{ij}^{a=1}=0)\right\}$.
The estimand (\ref{DEestimand}) is not a causal effect, as it comprises a comparison of different cluster-level outcomes, namely $Y_i^{a=1}$ and $\tilde{Y}_i^{a=1}$. As noted above, for an estimand to have a causal interpretation, the same outcome must be compared under different counterfactual scenarios.

It is conventional, although not incontrovertible\cite{Pearl2018}, to define causal effects only for a treatment or exposure that is manipulable, i.e., there can be ``no causation without manipulation''\cite{Holland1986}. If this convention is followed, then in cluster-randomized trials with non-participation, the direct effect of vaccination would only be considered well defined in always participators. Otherwise, to define the relevant potential outcomes would require considering a counterfactual scenario where non-participators receive vaccine. However, for the study design under consideration, always participators receive vaccine if and only if other always participators in their cluster also receive vaccine. Thus it is not possible to observe both (i) a vaccinated always participator and (ii) an unvaccinated always participator, while holding fixed the proportion of other individuals who are vaccinated in the cluster; hence the direct effect is not identifiable without additional assumptions.

On the other hand, if the ``no causation without manipulation'' convention is not adopted, there are other complications that may arise with estimating the direct effect. In particular, in cluster-randomized trials with non-participation, vaccine coverage within a cluster is dictated by the collective level of individual participation in the study, which is not under the investigator's control. Factors associated with participation may also be associated with the outcome of interest, creating the potential for confounding. Thus causal inference methods for observational studies, such as those assuming no unmeasured confounding, would in general be necessary to draw inference about direct effects, e.g., see Tchetgen Tchetgen and VanderWeele\cite{Tchetgen2012}, Perez-Heydrich et al.\cite{Perez-Heydrich2014}.

\section{Typhoid Vaccine Trial} \label{MotivatingExample}
A cluster-randomized study was conducted to investigate the effectiveness of a Vi polysaccharide typhoid vaccine in Kolkata, India over two years of follow-up from 2004 to 2006 \citep{Sur2009}. The control in this trial was an inactivated hepatitis A vaccine. Geographic mapping and a census that characterized and counted all people and households in the study area were used to define 80 clusters. For purposes of randomization, clusters were stratified by ward (an administrative unit of Kolkata) and by the number of residents in certain age groups. Overall, 40 clusters were assigned to Vi vaccine and the other 40 to control. Because data from the typhoid trial are not publicly available, a simulated data set was constructed (see Supporting Information). The data were simulated to match exactly the cluster level summary statistics from the actual trial shown in Table \ref{margins}.

\begin{center}
	\begin{table}[ht]%
		\centering
	\caption{Summary statistics of a cluster-randomized study in Kolkata from 2004 to 2006 of a Vi typhoid vaccine versus a hepatitis A control vaccine. \cite{Sur2009}}
	\label{margins}
	\begin{tabular*}{500pt}{@{\extracolsep\fill}lcccc@{\extracolsep\fill}}
		\toprule
		& \textbf{Typhoid Vaccine} & \textbf{Control}\\
	\midrule
		Number of clusters & 40 & 40\\
		Mean $\pm$ SD* of people per cluster & 777 $\pm$ 136 & 792 $\pm$ 142\\
		Mean $\pm$ SD of participants per cluster & 472 $\pm$ 103 & 470 $\pm$ 104\\
		Number of participants & 18869 & 18804\\
		Number of non-participants & 12206 & 12877\\
		Number of events in participants & 34 & 96\\
		Number of events in non-participants & 16 & 31\\
		\bottomrule
\end{tabular*}
\begin{tablenotes}
	\item[*] *Standard deviation.
\end{tablenotes}
\end{table}
\end{center}

Sur et al.\cite{Sur2009} measure vaccine effects in terms of hazard ratios. However, causal interpretations for hazard ratios are difficult because hazard ratios can depend on time and have an inherent selection bias \citep{Hernan2010}. In particular, time-specific hazard ratios compare different subsets of subjects and, as noted above, estimands have a causal interpretation only when comparing potential outcomes between the same set (or subset) of units. Due to these issues, instead of using the hazard ratio to determine the vaccine effects as in Sur et al.\cite{Sur2009}, the risk difference of typhoid over two years is calculated here to quantify vaccine effects.

The overall, indirect, and total effects were estimated using (\ref{eqestimator}) with the $Y_i$ definitions provided in section \ref{OE}. The effect estimates, estimated standard errors (SEs), and 95\% Wald confidence intervals (CIs) are shown in Table \ref{estimates}. On average, over the study period of two years, Vi clusters had 1.61 cases of typhoid per 1000 people, while control clusters had 4.10 cases of typhoid per 1000 people. The overall effect estimate is -2.49 cases per 1000 people with 95\% CI (-3.41, -1.58). The overall effect estimate has a straightforward interpretation which may be of interest to public health officials such as epidemiologists. In particular, the number of cases of typhoid per 1000 persons over a two year period is estimated to decrease by 2.5 on average when a cluster receives the Vi vaccine compared to receiving control.

Both participants and non-participants appear to benefit from the Vi vaccine. In particular, the total effect estimate is -3.30 (95\% CI -4.61, -1.99), indicating that assigning a cluster to Vi vaccine causes 3.3 fewer cases of typhoid per 1000 participants compared to assigning a cluster to hepatitis A vaccine. Likewise, the indirect effect estimate suggests that assigning a cluster to the typhoid vaccine results in 1.29 (95\% CI 0.19, 2.38) fewer cases per 1000 non-participants; as non-participants never receive the vaccine, this indicates an indirect (or herd immunity) effect of the typhoid vaccine.

\begin{center}
	\begin{table}[hh!]%
		\centering
	\caption{Estimates of overall, indirect, and total effects, standard errors (SE), and 95\% Wald confidence intervals (CI). Effect estimates are differences in typhoid cases per 1000 people per two years.}
	\label{estimates}
\begin{tabular*}{500pt}{@{\extracolsep\fill}lcccc@{\extracolsep\fill}}
	\toprule
\textbf{Effect} & \textbf{Estimate (SE)} & \textbf{95\% CI}\\
	\midrule
Overall & -2.49 (0.47) & (-3.41, -1.58)\\
Indirect & -1.29 (0.56) & (-2.38, -0.19)\\
Total & -3.30 (0.67) & (-4.61, -1.99)\\ 
	\bottomrule
\end{tabular*}

\end{table}
\end{center}

On the other hand, the naive direct effect estimator (\ref{DEestimator}) equals 0.56 (95\% CI -0.44, 1.55).  Although not statistically significant, this point estimate implies that the average number of cases of typhoid per 1000 people is higher in vaccinated individuals compared to non-vaccinated individuals in clusters randomized to the Vi vaccine. However, as described above, this estimate cannot be interpreted as an effect of the vaccine because it does not account for possible confounding. For example, perhaps individuals at higher risk of typhoid chose to participate in the trial, or those who participated tended to have different health care seeking behavior. Moreover, the average number of cases of typhoid per 1000 people was also higher in participants compared to non-participants (2.57, 95\% CI 1.19, 3.96) in the control clusters, providing direct evidence of confounding. Sur et al.\cite{Sur2009} reported similar results, with incidence of typhoid higher in participants compared to non-participants, both within Vi vaccine clusters and within control clusters. 


\section{Discussion} \label{Conclusion}
Randomized controlled trials are the gold standard in vaccine trials since randomization ensures that the vaccine and control groups are comparable. Carefully defining estimands in clinical trials is vital to ensure accurate interpretation of the resulting treatment effect estimates. Because cluster-randomized trials can be large and expensive to conduct, it is important to formally characterize estimands for use in these trials. This paper considers causal estimands in cluster-randomized trials where interference may be present within clusters. An illustrative example is provided motivated by a recent cluster-randomized typhoid vaccine trial demonstrating inference and interpretation of the overall, total, and indirect effect estimands. These types of analyses can be used to inform public health policies regarding vaccination.


In cluster-randomized trials with self-selection, estimators of the direct effect must account for possible confounding. A standard method to adjust for confounding is to condition on covariates and assume that conditional on these covariates, participants and non-participants are exchangeable. A possible indirect way to adjust for confounding could involve comparing outcomes between participants and non-participants in the control clusters as an estimate of the confounding bias, if present, similar to negative control approaches described in Lipsitch et al.\cite{Lipsitch2010} and Tchetgen Tchetgen\cite{Tchetgen2013}. Alternatively, two-stage randomized designs could be considered to eliminate possible confounding when drawing inference about the direct effect. In two-stage randomized experiments, clusters are first randomly assigned to a treatment allocation program, then individuals within those clusters are assigned to treatment or control based on their cluster's treatment allocation program \cite{Hudgens2008}. Randomization eliminates possible confounding at the cluster and individual level, such that direct, indirect, total, and overall effects can be estimated \cite{Hudgens2008, Baird, Basse2017}. However, it may not always be feasible to conduct two-stage randomized trials. In addition, the effects estimated by a two-stage randomized experiment are not equivalent to the effects estimated in cluster-randomized trials with participation self-selection and may have less public health relevance \cite{Barkley2017, Papadogeorgou2017}.


Estimated effects may have greater real-world relevance depending on the estimands of interest and characteristics of individuals in the trials, such as the level of participation. Westreich\cite{Westreich2017} provides several examples of population intervention effects defined by contrasts in average potential outcomes under different possible interventions on the distribution of treatment. These population intervention effects may be more germane to real-world policy than the traditional approach of defining causal effects by comparing average outcomes when all individuals in the population receive treatment versus when no individuals receive treatment. The estimands described here for cluster-randomized trials with self-selection are examples of population intervention effects, to the extent that the participation rate in the trial approximates vaccination uptake should the vaccine under evaluation become widely available to the public. For example, in Sur et al.\cite{Sur2009}, about 60\% of individuals on average chose to be vaccinated in both Vi and hepatitis A clusters; thus, the overall, total, and indirect effect estimates approximate the effects of vaccinating 60\% of the population. Such effect estimates could potentially help inform public health policy decisions regarding vaccination.

\section*{Acknowledgments}
The authors thank the Causal Inference with Interference working group in the Biostatistics Department at UNC-Chapel Hill: Shaina Alexandria, Brian G. Barkley, Bryan Blette, Sujatro Chakladar, Bradley Saul, and Bonnie Shook-Sa for their helpful suggestions. This work was partially supported by NIH grants R01 AI085073 and T32 ES007018.


\bibliography{ClusterRandCausalUpdated}

\end{document}